\documentclass[times,final]{elsarticle}
\biboptions{numbers,sort&compress}

\usepackage{journalarx}
\usepackage{framed,multirow}

\usepackage{amssymb}
\usepackage{latexsym}
\usepackage{xcolor}
\usepackage[colorlinks=true,urlcolor=blue,citecolor=blue,linkcolor=blue]{hyperref}
\usepackage{uri}
\definecolor{newcolor}{rgb}{.8,.349,.1}

\newcommand{\figref}[1]{Fig.~\ref{#1}}

\usepackage{siunitx}
\DeclareSIUnit\gauss{G}

\newcommand{\rhl}[1]{\textcolor{black}{#1}}
\newcommand{\hl}[1]{\textcolor{black}{#1}}

\usepackage{xfp}
\usepackage[labelfont=bf,justification=raggedright,singlelinecheck=false,textfont=normalfont]{caption}
\makeatletter
\newcommand{\ORCID}[1]{\hspace{\fpeval{\f@size*0.25} pt}\raisebox{-\fpeval{\f@size*0.15} pt}{\href{https://orcid.org/#1}{\includegraphics[height=\f@size pt]{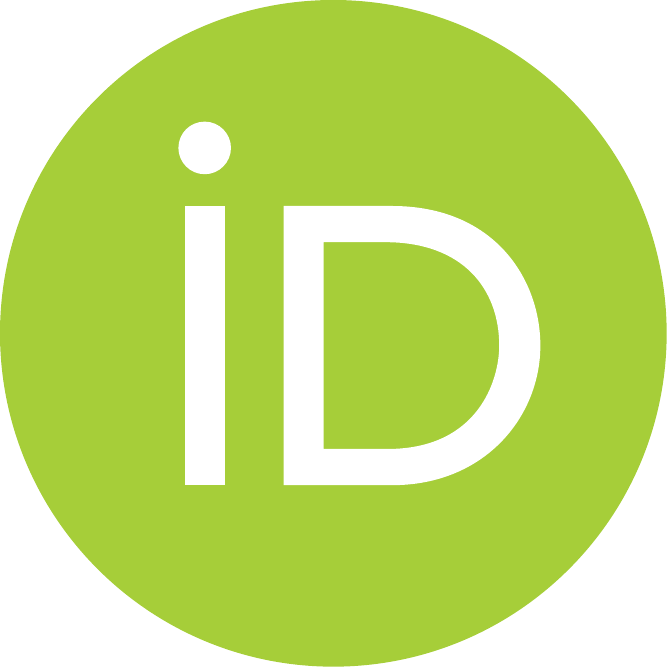}}}\hspace{\fpeval{\f@size*0.25} pt}}
\makeatother

\begin{document}
\verso{M. Chilcott, \textit{et. al.}}
\begin{frontmatter}
  \title{{\large Tutorial}\\Low-cost Wireless Condition Monitoring System for an Ultracold Atom Machine}
  % \tnotetext[mytitlenote]{Fully documented templates are available in the elsarticle package on \href{http://www.ctan.org/tex-archive/macros/latex/contrib/elsarticle}{CTAN}.}
  
  \author{Matthew Chilcott\ORCID{0000-0002-1664-6477}}
  % \author{N. Kj{\ae}rgaard\corref{mycorrespondingauthor}}
  \author{Niels Kj{\ae}rgaard\ORCID{0000-0002-7830-9468}}
  \address{Department of Physics, QSO---Centre for Quantum Science, and Dodd-Walls Centre for Photonic and Quantum Technologies, University of Otago, Dunedin, New Zealand}
  
  \received{\today}
  % \finalform{10 May 2013}
  % \accepted{13 May 2013}
  % \availableonline{15 May 2013}
  % \communicated{Name}

  \begin{abstract}
    We present a flexible wireless monitoring system for
    condition-based maintenance and diagnostics tailored for dynamic
    and complex experimental setups encountered in modern research
    laboratories. Our platform leverages an Internet-of-Things
    approach to monitor a wide range of physical parameters via
    wireless sensor modules that broadcast to a networked computer. We
    give a specific demonstration for a so-called ultracold atom
    machine, which is the workhorse of many emerging quantum
    technologies and \hl{marries a broad spectrum of equipment and
      instrumentation into its setup. As a distinctive feature, our
      monitoring system taps into physical parameters of the ultracold
      atom machine both via customized sensor modules that directly
      perform measurements, and via modules that recruit and annex
      constituent instruments of the machine} \rhl{for the additional
      purpose} \hl{of retrieving information for diagnostics.}
    
    \vspace{4em}
    %%%% 
  \end{abstract}
  
  \begin{keyword}
    % \MSC 41A05\sep 41A10\sep 65D05\sep 65D17
    %% Keywords
    \KWD Internet of things\sep Ultracold Atom Machine\sep Condition Monitoring\sep Process Monitoring\sep Sensor Networks\sep Wireless
    Instrument Control, Wireless Sensors
  \end{keyword}
  
\end{frontmatter}

\section{Introduction}

\begin{figure}[h]
	\centering
	\includegraphics[width=0.9\columnwidth]{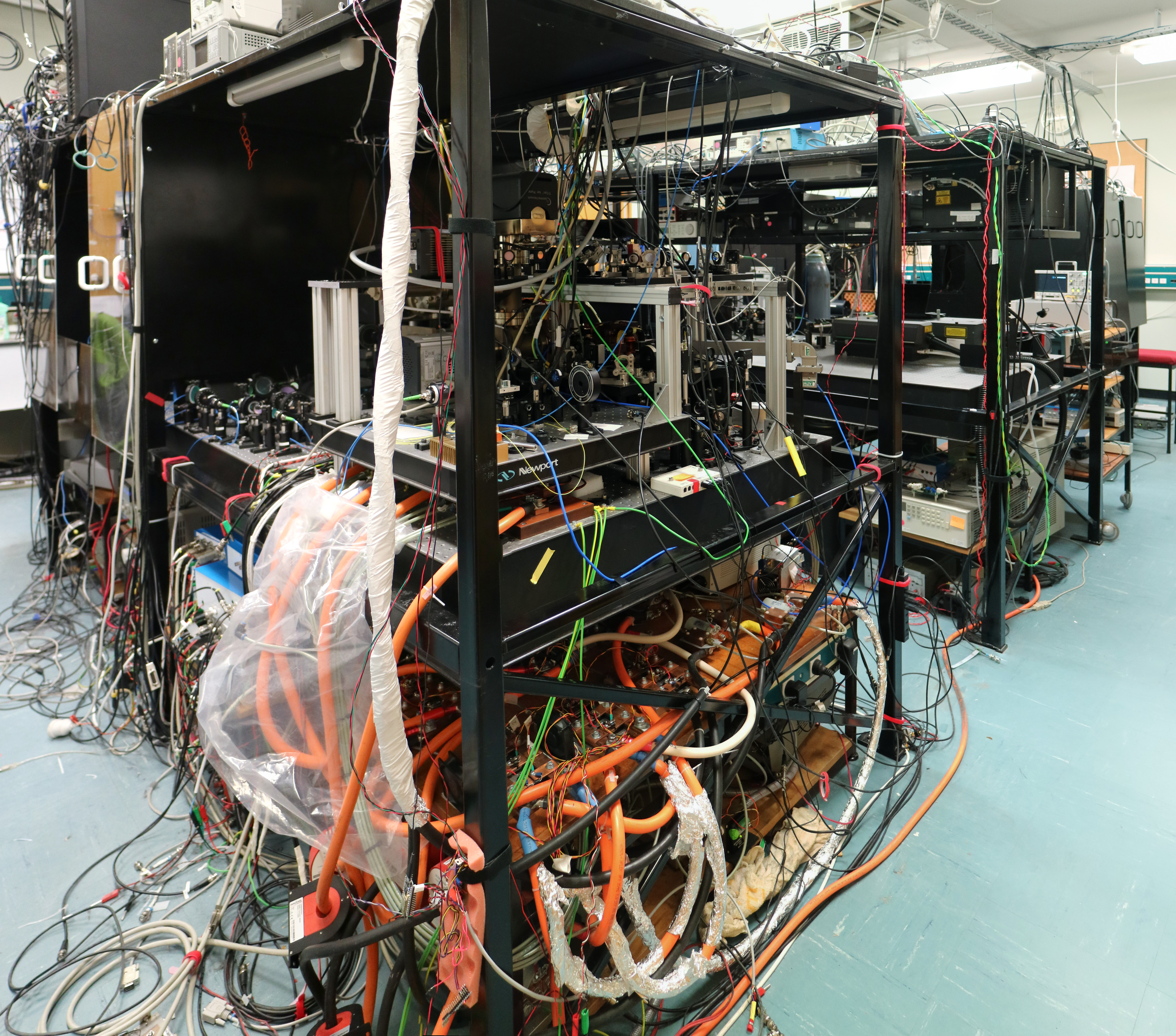}
	\caption{\label{fig:labphoto}\rhl{The ultracold atom machine
            at the University of Otago.  Equipment spans three \SI{1.2
              x 3}{\metre} optical tables in four tiers. The experiment confines
            atoms inside an ultra-high vaccum chamber using
            high-current-carrying ($\sim 400$~A) magnetic field
            coils. The trapped atoms are manipulated by $\sim 10$
            different laser systems, 3 different microwave and RF
            emitters, and $\sim 10$ auxiliary magnetic field coils. The
            machine is a hybrid of many different technologies from
            optical, microwave, vacuum, electronic, and mechanical
            engineering and incorporates a range of both commercially
            available equipment and custom-built hardware.}}
\end{figure}

The advent of laser cooling opened up a way to produce trapped atomic
gases at temperatures only a few microkelvin above absolute zero,
initiating the field of ultracold atomic physics
\cite{Fallani2015}. Augmenting laser cooling with evaporative cooling,
a push was made to the nanokelvin domain where Bose-Einstein
condensates and degenerate Fermi gases were attained. These ultracold
atomic systems have remained at the cutting-edge of research because
they offer a pristine environment to explore fundamental quantum
phenomena, and the experimental platforms used in their production can
accurately be referred to as `ultracold atom machines'
\cite{Streed2006, Lewandowski2003}. Such machines\rhl{---an example is shown in \figref{fig:labphoto}---}are extremely
complex, hybridising many different technologies from optical,
microwave, vacuum, electronic, and mechanical engineering. They
typically incorporate a range of both commercially available equipment
and custom-built hardware, patching a variety of systems together to
get a functional ultracold atom machine and an optimised process
sequence.  The end product of running a process cycle---a small gas
cloud a few billionths of a degree above absolute zero temperature,
levitated inside a vacuum chamber by electromagnetic fields---is
extremely sensitive to changes in the process parameters and the
environment.  A reproducible result is hence strongly tied to these
conditions remaining stable. Accordingly, detected variations in the
end product such as atom number, quantum state purity or sample
temperature may be associated with variations in specific, monitored
process parameters through suitable analysis.  For example,
Ref. \cite{krarup} employed commercial data acquisition hardware to
monitor 50 parameters of an ultracold atom machine and tied
fluctuations in the final atom number to the temperature of a magnetic
field coil.

Condition monitoring \cite{Vachtsevanos2006} is a well known paradigm
in conventional industrial plants, where monitoring equipment parameters
allows for condition-based or predictive maintenance and leads to
increased equipment up-time and smaller maintenance and operating
costs. These benefits could also be reaped in a research laboratory
setting, where, however, particular challenges emerge when dealing with
commercial test and measurement equipment. Past generations of, for
example, microwave frequency synthesizers, current supplies, and laser
controllers will not natively interface with a condition monitoring
system. Such equipment might however represent a considerable
investment and from performance point of view remain completely
adequate: a high-end Hewlett Packard 26 GHz YIG based synthesizer from
the 1980s remains a fine instrument today. Incorporation of such
instruments into a monitoring system along with a range of sensor
modules can be achieved using `Internet-of-Things' (IoT) technologies
and ideas -- an approach which is formed around having large numbers
of devices (`things') connected to the Internet \cite{IoT}, and which has
grown in popularity with the availability and declining cost of
hardware with networking capability. IoT ideas are also finding use in
home automation \cite{SUNG2020106997}, wearable technology, healthcare
\cite{10.1371/journal.pmed.1001953, Akka2020, Sokullu2020}, and are moving into wireless
sensor network technologies in conventional processing plants
\cite{Industry4, IEEE_ConditionMonitoring}. Wireless sensor networks
themselves are used in numerous settings including sailboats
\cite{IEEE_Sailing}, launch vehicles \cite{IEEE_LaunchVehicle},
heritage buildings \cite{IEEE_Heritage},
agriculture\cite{AgriculralWSN,Putra2020}, composting
\cite{Casas2014}, and environmental \cite{IEEE_Chloride, OrozcoLugo2020} and land
slide monitoring \cite{Khoa2018,IEEE_Landslide}.

\begin{figure*}[t]
  \centering
  \includegraphics[width=0.8\linewidth]{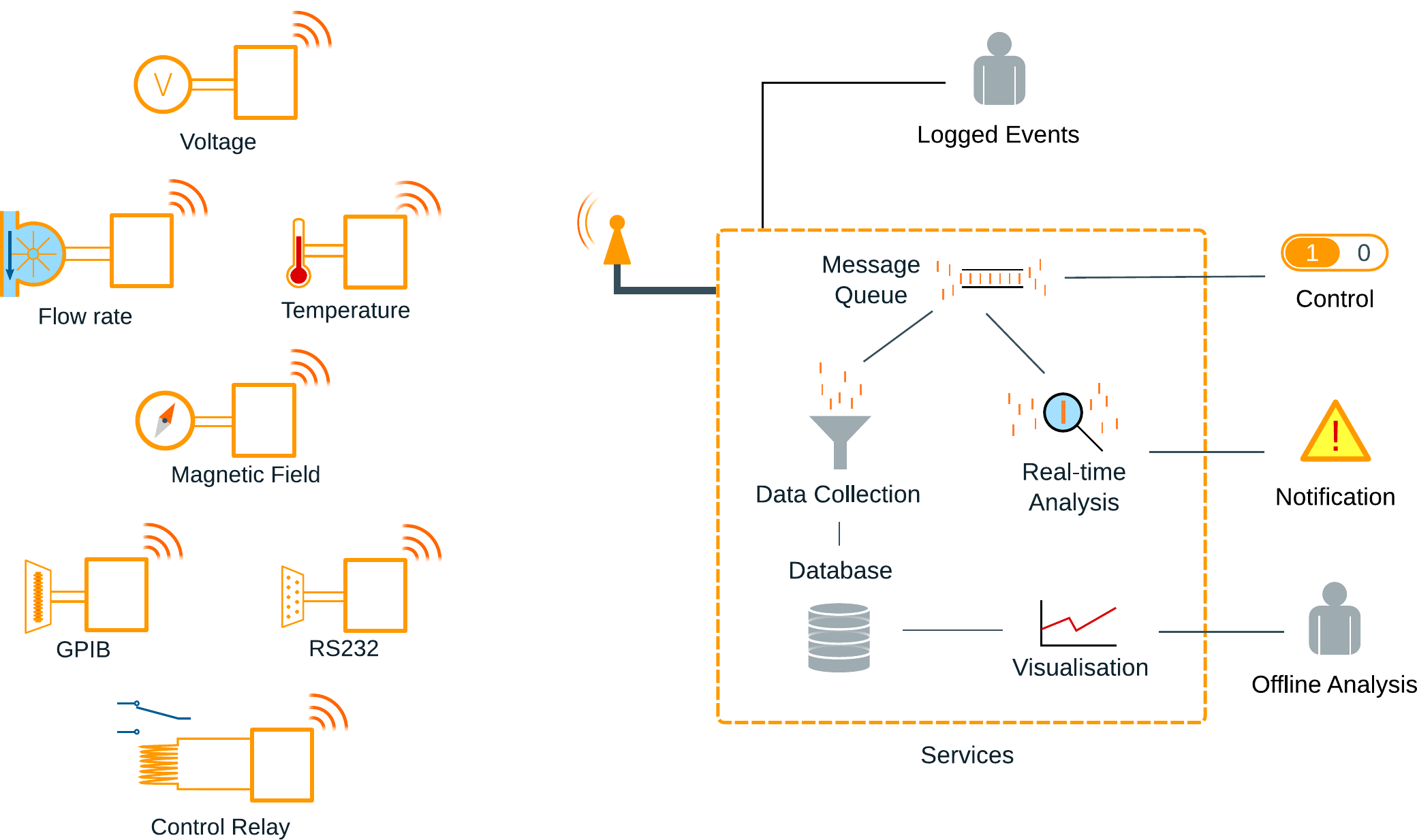}
  \caption{\label{fig:system}The architecture and data-flow pipeline of
    the monitoring system. Some examples of the sensors we use are
    shown.}
\end{figure*}

In this paper, we focus on a wireless, low-cost, long-term monitoring
system, with an emphasis on experimental apparatus, which \hl{entails
  interfacing with both commercial laboratory equipment (for example
  via IEEE-488 or RS-232) and home-built, customized sensors using
  IEEE 802.11b/g/n WiFi as the communication backbone. Our monitoring
  system straddles these dual domains} and augments a wireless sensor
network with services for data collection and analysis as illustrated
in \figref{fig:system}. We have tested our monitoring system over a
period of 2 years in connection to an ultracold atom machine at the
University of Otago which over the past decade has developed into an
apparatus of significant complexity
\cite{Rakonjac2012,Deb2013,Deb2014,Chisholm2018,Thomas2018,Sawyer2019}. We
have found the operation of the monitoring system robust,
reliable\footnote{While we have not performed formal reliability tests \cite{Moore2020}, during two years of continuously running the system, we registered only four instances of unscheduled interruption of data collection. In all cases the fault could be traced to the commercial wireless access point (WAP) which after a reboot would restore the system. After identifying the WAP as a weak link, we replaced it with more modern hardware from a different vendor, and we expect increased up-time in the future.}, and useful in detecting equipment failures in
our laboratory, especially those of a chiller used for cooling water,
and the lab air conditioning -- saving us from further equipment
failure, and answering the questions ``does it feel warm in here?''
and ``why is the experiment performing badly'' with logged data. More
recently, we have extended our data logging framework to contact
tracing of lab users during the COVID19 pandemic. In the following we
will describe key features of our monitoring system, including the
construction of wireless sensor modules, examples of their usage and
their integration into a flexible architecture that can log data from
a diverse range of sources.\footnote{Reference \cite{OurProject}
  provides ready-to-use resources (sensor module firmware, service
  automation, and PCB designs for manufacture) and gives detailed
  documentation on how to replicate our monitoring system.}

\section{Sensor Module}
Our sensor modules are based on either the ESP8266 or ESP32
micro-controllers by Espressif Systems which are self-contained
System-on-Chips with built-in WiFi. These modules also contain
analog-to-digital converters (ADCs), digital serial capabilities
(including UART, I2C, SDIO, SPI), and a number of general-purpose
input-output (GPIO) pins. We specifically use the NodeMCU V1.0
\cite{NodeHW} and the DevKit C development boards for the ESP8266 and
ESP32, respectively, which add USB serial interfaces for convenient
programming as well as a supply voltage regulator.

Each of our sensor modules is programmed with firmware we have
developed\footnotemark[\value{footnote}] according to the application
and containing drivers for the connected sensor. Table~\ref{tbl:bom}
shows the cost of the development boards, and the extra hardware for
three typical example nodes---a simple temperature sensor, an RS-232
interface, and a wireless General Purpose Interface Bus (GPIB) module.

Our code \cite{OurProject} provides a ready-to-use framework where
firmware is uploaded to the ESP development board via its USB connection (we
use the PlatformIO tools \cite{platformio} for this purpose). Once
deployed, the device can be reprogrammed via WiFi from a
web-browser. Our code library contains detailed information and
instructions on configuring and deploying the modules. Essentially,
the steps to be followed can be summarized as:
\begin{itemize}
\item Set up the data collection services on a networked computer in the laboratory (only needs to be done once).
\item Connect sensors to the ESP development board.
\item Specify the connected sensors in the code.
\item Build firmware on computer (via PlatformIO) and upload it to the
  development board via USB (the sensor module will then broadcast a WiFi network to configure its connection).
\item Deploy sensor module in lab.
\item Connect to WiFi broadcast by the sensor module to connect to WiFi in the lab and publish measurements to the data collection server.
\end{itemize}

Below we present a range of the sensors that
we have deployed in our laboratory and for which we provide code and hardware examples.

\begin{table}
  \centering
  \renewcommand{\arraystretch}{1.1}
  \caption{\label{tbl:bom}Typical Bill of Materials for monitoring nodes.}
  %\resizebox{0.9\linewidth}{!}{
    \begin{tabular}{| l | r |}
      \hline
      Part & Price (USD) \\
      \hline
      \multicolumn{2}{|c|}{Controller Development Boards}\\\hline
      ESP32 DevKit C  &\$7.49\\
      ESP8266 (NodeMCU V2)  &\$5.99\\
      \hline
      \multicolumn{2}{|c|}{GPIB Monitor}\\\hline
      GPIB Connector (e.g. DigiKey 1024RMA-ND) & \$5.53\\
      Arduino Nano  & \$4.66 \\
      \SI{1}{\kilo\ohm} Resistor & \$0.01\\
      PCB Manufacture & \$2.00\\
      \hline
      \multicolumn{2}{|c|}{RS232 Monitor}\\\hline
      DB9 Connector  & \$0.77\\
      MAX202 Transceiver & \$1.03 \\
      \SI{1}{\kilo\ohm} Resistor & \$0.01\\
      $4 \times$\SI{100}{\nano\farad} Capacitors & \$0.01\\
      PCB Manufacture& \$2.00\\
      \hline
      \multicolumn{2}{|c|}{Temperature Monitor}\\\hline
      DS18B20 Thermometer  & \$1.69\\
      \hline
    \end{tabular}
   % }
\end{table}

\subsection{Temperature Sensor}
For contact temperature measurement we primarily use the DS18B20 from
Maxim Integrated, which we fit into various places in our machine.
Each sensor has a digital interface, and is capable of
reporting the temperature to a resolution as good as
\SI{62.5}{\milli\kelvin} \cite{DS18B20_datasheet}.  We find the TO-92
package (small ``transistor'' shape) of the DS18B20 to be a convenient
size to locate throughout our apparatus. We also employ variants
enclosed in a water-proof housing.

\begin{figure}[b!]
  \centering
  \includegraphics[width=\linewidth]{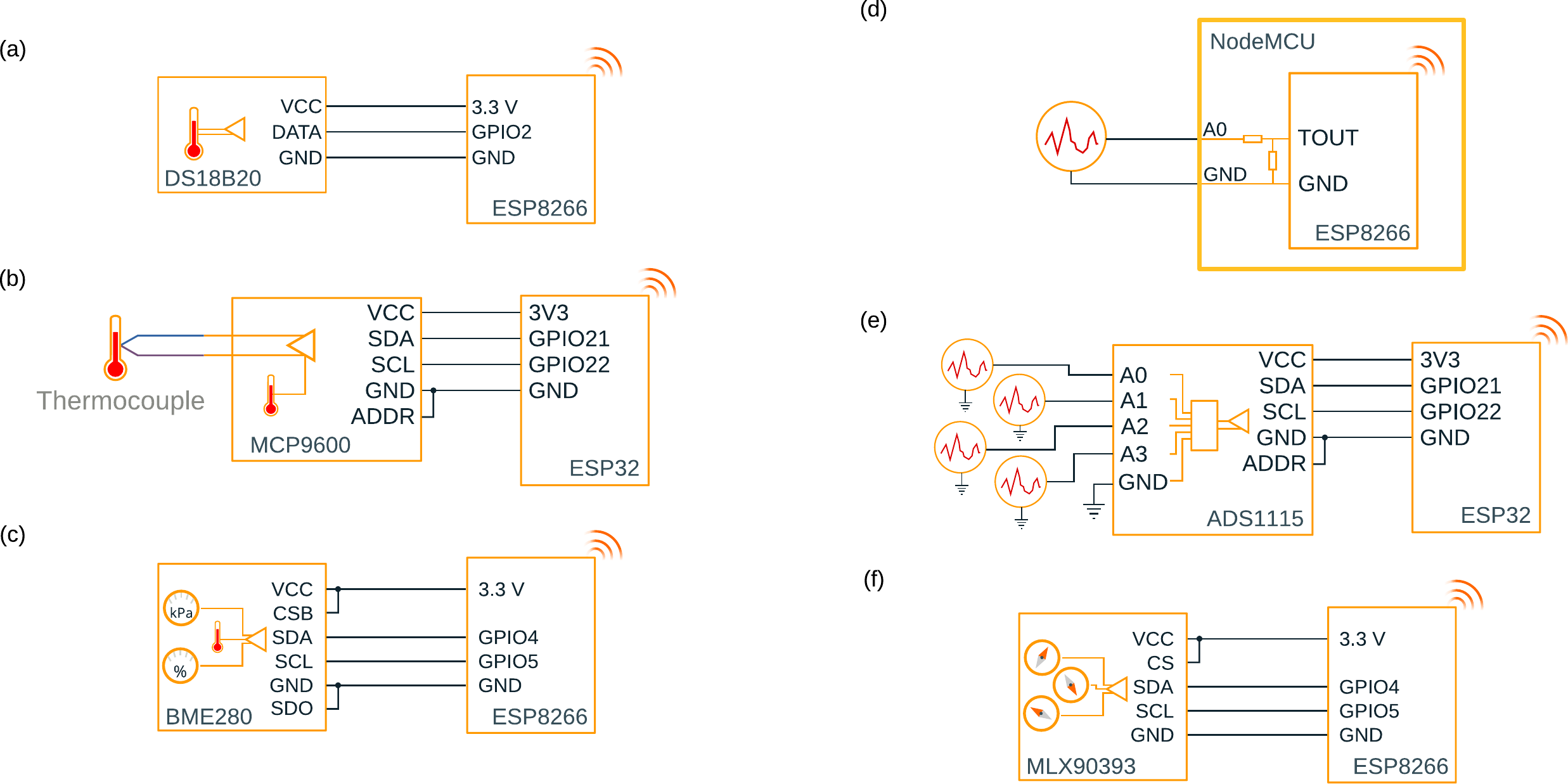}
  \caption{\label{fig:wiring_example}Connection diagrams for wireless
    sensor modules for measuring contact temperature (a), (b),
    environmental temperature humidity and pressure (c), analog
    voltages (d), (e), and magnetic fields (f). The modules (b), (c),
    (e), and (f) integrate I2C sensors, which rely on power lines and
    two connections for the serial interface. Many of the sensors we use
    additionally require configuration by setting the voltage of
    another pin on the sensor, e.g. CS, CSB, SDO and ADDR above}
\end{figure}

To construct a wireless sensor module from the DS18B20, one can simply
connect it to an ESP8266 as shown in \figref{fig:wiring_example}(a), then
configure the firmware specifying a DS18B20 is connected, and a handle
to identify the measurement. Multiple DS18B20 devices can be connected
to the same GPIO pin and read out separately as they have unique IDs.

For an even smaller sensor head or to measure temperatures exceeding
\SI{125}{\celsius}, we use K-type thermocouples, with the MCP9600
16-bit thermocouple-to-digital converter from Microchip (see
Fig.~\ref{fig:wiring_example}(b)). This device features all the
required signal conditioning, supports multiple types of
thermocouples, and has built-in compensation for the cold junction
formed by physically connecting the thermocouple to the converter.

\subsection{Atmospheric Sensors}
To measure atmospheric signals of interest, such as ambient
temperature and humidity, we use the AM2302 from Aosong Electronics --
a small digital sensor which can measure the ambient temperature and
relative humidity. Another sensor we use for this purpose is the
BME280 from Bosch Sensortec, which additionally measures the
atmospheric pressure (see Fig.~\ref{fig:wiring_example}(c)).

From these signals one can, for example, detect changes in the lab
environmental control or the function of a positive pressure
environment. In particular, we have found it useful to monitor the
outside temperature with an atmospheric sensor module mounted on the
side of the building to examine the impact of weather on our cooling
water system's performance. Monitoring the performance of the lab air
conditioning has also proven useful.

\subsection{Analog Voltage Measurements}

Virtually any physical quantity we want to measure can be converted to
an analog voltage by some electrical transducer, for example:
currents (using a sense resistor, or hall-effect sensor), resistance
(using a bridge circuit), temperatures (using a thermistor),
forces/pressures (with a strain gauge), and magnetic fields (with
hall-effect sensors). Furthermore, many industrial sensors produce an
analog signal of 4--20~\si{\milli\ampere}, and many research
instruments provide a voltage output to monitor signals of
interest. The variety of applications for monitoring analog signals
makes this an important  requirement of our framework.
\figref{fig:wiring_example} (d) and (e) shows two configurations for this task.

\subsubsection{Internal ADC}
The ESP8266 features an \rhl{on-chip} 10-bit ADC with a native
\SIrange{0}{1}{\volt} range, which on the NodeMCU boards has been
extended to a \SIrange{0}{3.2}{\volt} range with a
{220:100}~\si{\kilo\ohm} resistive voltage divider. This range can be
adjusted by suitable replacement of the resistors of the NodeMCU board
or by adding a resistor inline with the signal.

The ESP32 includes two 12-bit ADCs, which can be configured to take
maximum voltages of up to \SI{3.3}{\volt}. These ADCs can be
configured to measure on a number of different GPIO pins, meaning that
one can monitor multiple signals in parallel. However, the ESP32's
ADCs suffer from non-linearity in their readings \rhl{(see
  \hyperref[app]{supplementary material})}, so we recommend using the
ADS1115 external ADC, which we describe below.

The \rhl{on-chip} ADC is referenced to the chip ground, and hence the
ESP controller must share its ground with the signal. The ground does, however, not have
to be at the earth potential, as the controller can be powered from an
isolated/floating power supply or battery. This allows for performing
measurements in galvanic isolation from earthed equipment.

Our testing \rhl{(see \hyperref[app]{supplementary material})} has
demonstrated that the analog input pins of the ESP controllers are not
tolerant of negative voltages (less than \SI{-0.3}{\volt}), or
voltages exceeding \SI{3.3}{\volt}. It should be noted that the
controller should be powered whenever it is connected to a signal
source, as the input's behaviour and tolerance changes when not
powered.

The \rhl{on-chip} ADC can be sampled at a rate of about
\SI{2}{\kilo\hertz}, but for most of our monitoring applications, we sample at much
lower rates, e.g., \SI{1}{\hertz}.

\subsubsection{External ADC}
For higher resolution measurements, we use the ADS1115 analog-digital
converter from Texas Instruments, which provides higher-resolution (16
bit) conversion, on four input channels, and a programmable-gain
input. This allows a maximum measurement range of
\SI{\pm6.144}{\volt}, though the input pins can only tolerate
\SIrange{-0.3}{5.3}{\volt} when run off a \SI{5}{volt}
supply. Negative signals are only possible with differential
measurements between two positive voltages. Similarly to the ESP's internal
ADC, one can make floating measurements with care.

\subsection{Magnetic-field Sensors}

Ultracold atoms are particularly sensitive to magnetic
fields. Hall-effect sensors are one way of measuring these fields and
present an analog signal which can be monitored as previously
described. We use a digital magnetic field sensor, the MLX90393 from
Melexis. This is a 16-bit, 3-axis magnetometer capable of measuring
fields of up to \SI{\pm 500}{\gauss}, which we find suitable for
monitoring our magnetic traps and background fields. A variety of
alternative digital 3-axis magnetometers exist, but many are
manufactured as digital compasses and saturate at fields of as little
as \SI{4}{\gauss}, which is insufficient for applications near our
electromagnets. A wireless module based on the MLX90393 sensor is
depicted in \figref{fig:wiring_example}(f).

\subsection{Digital Signal Monitoring}
Commercial instruments embedded in our ultracold atom machine often
have TTL outputs, e.g. ``PLL locked'', ``Laser locked'' or ``Error'',
which we track via sensor modules. Other interesting digital signals
include laser interlock status, comparator outputs, or timing signals,
so the monitoring system is aware of the state of a process cycle.

The GPIO pins of the ESP controllers can be used to monitor the state
of digital signals. The controllers are \SI{3.3}{\volt} devices and
the ESP32's GPIO pins will not tolerate higher voltages, but the
ESP8266's GPIO pins are \SI{5}{\volt} tolerant and can accept standard
\SI{5}{\volt} TTL signals \rhl{(see \hyperref[app]{supplementary
    material})}. For higher voltages, a resistor should be put in
series with the signal to limit the current to less than
\SI{12}{\milli\ampere}, preventing damage of the GPIO pins.

\subsubsection{Time-sensitive Digital Signals}
Our flow meters for cooling water produce pulsed signals with the
pulse rate being a measure of the flow rate. The GPIO pins of the ESP
controller could be used for these tasks, but the processor in the ESP
controller is busy managing WiFi tasks, and running a real-time
operating system, which is not ideal for monitoring time-sensitive
signals. The ESP32, however, has an \rhl{on-chip} pulse counter peripheral
which can be used. For the ESP8266 we have used an external processor
(an ATmega328P from Microchip on an Arduino Nano 3 board
\cite{ArduinoNano}) to perform time-critical functions. The ATmega328P
has a convenient timer-counter peripheral for this purpose.

To measure the pulse rate of our flow meters we have one of these
external boards programmed as a frequency counter and we read the
frequency out with the ESP controller via a digital serial connection
(I2C). The external processor can be used to make pulse width
measurements (e.g., monitor a laser beam shutter's opening time to
detect if it is stuck) or perform a Fourier transform on analog
signals to examine the spectrum of an incoming signal without loading
ESP controller.

\subsection{Serial Interface Readout}
\label{sec:serial}

\begin{figure}
  \centering
  \includegraphics[width=0.65\linewidth]{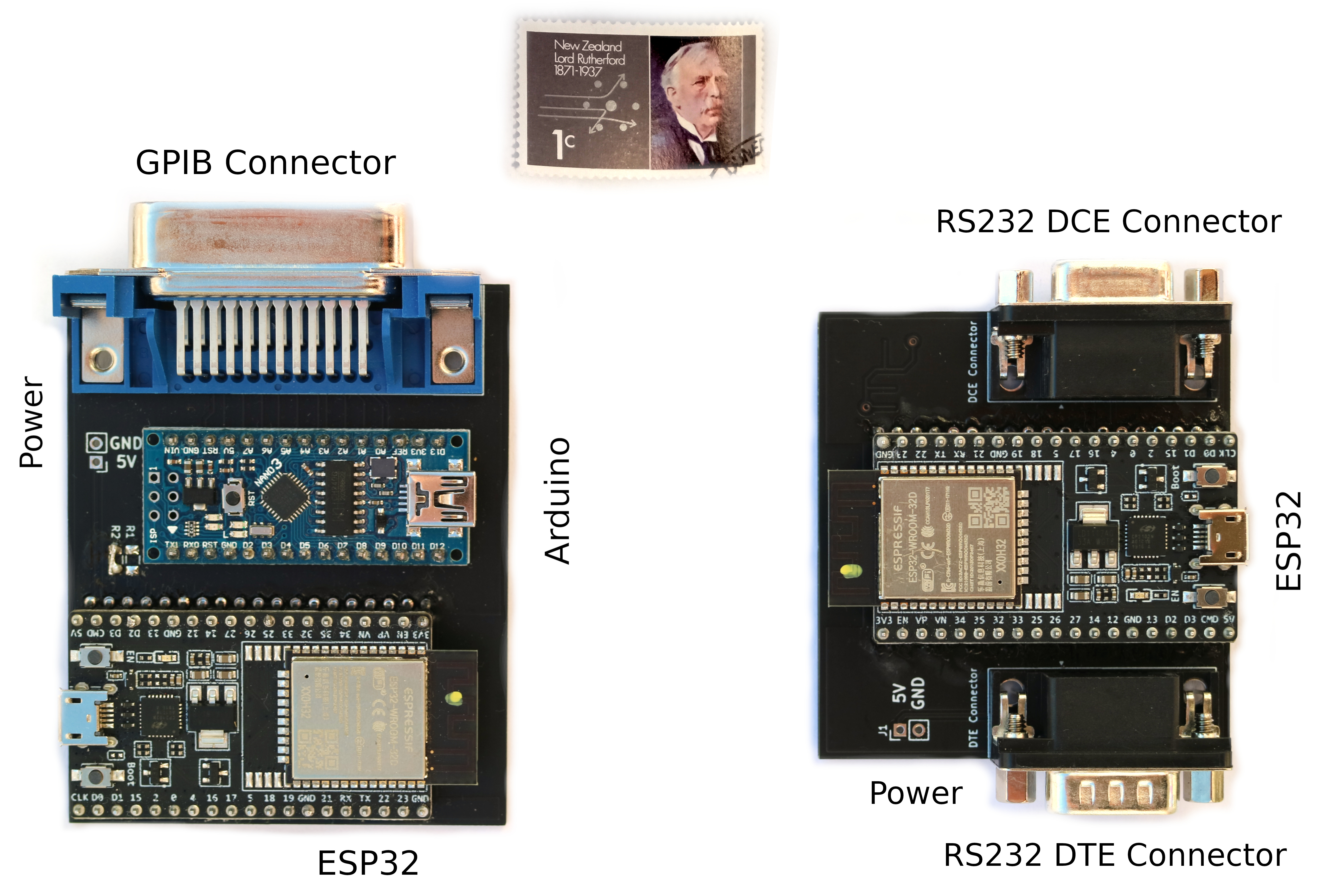}
  \caption{\label{fig:photo}{Left}: Our wireless GPIB interface
    which, for example, we use to monitor the power supplies of
    our electromagnets. The ESP32 uses an Arduino Nano to interface
    with GPIB. {Right}: Our wireless RS232 interface which, for
    example, we use to monitor a commercial laser controller. The module acts as Data Terminal Equipment (DTE), but many
    scientific instruments are wired as DTE, so it also features a
    Data Communications Equipment (DCE) wired connector to avoid the
    need for a null-modem cable. The MAX202 used to drive these serial
    interfaces is located on the reverse side of the board. A postage stamp is
    shown for size reference.}
\end{figure}

Commercial instruments are often outfitted with a digital serial
interface for remote control and diagnostics. These interfaces,
e.g. RS232, RS485 or GPIB, generally provide the ability to query the
status of the device (e.g. via SCPI). \hl{Despite the obvious usefulness
of wireless GPIB/serial interfaces} \cite{AR488, GPIBThesis},
\hl{curiously, commercial vendors do not provide solutions with integrated
wireless capabilities}, but invariably require a wireless router on top
of an already costly GPIB-to-Ethernet interface
\cite{NIWirelessGPIB, KeysightGPIB}. The cost of the latter can range from \$200 to
\$1600 depending on supplier. \hl{As an attractive and more compact
alternative, we provide a solution which integrates GPIB-to-WiFi on a
single printed circuit board (PCB)}, with a total component cost that is less than \$20.

To connect the ESP controller to the communication port of a
commercial instrument, we need a protocol translator capable of
handling and generating the voltage range of the interface bus. GPIB
or IEEE-488 is a parallel bus interface, which only requires TTL
signal levels. To manage this bus, we add an ATmega328P on an Arduino
Nano 3 \cite{ArduinoNano} board, following a design provided by
Ref. \cite{AR488}. This extra component acts as a level-shifter, and a
serial-to-parallel converter, making the system GPIB compatible. We
avoid using more powerful line drivers, which would be required for
driving a GPIB bus with many instruments---our largest single bus has
3 instruments---and leave our setup ``GPIB compatible'', rather than
fully compliant with relevant GPIB specifications. The ESP controller
and Arduino are assembled on a custom
PCB\footnote{\label{notey}Reference \cite{OurProject} includes
  production ready computer-aided manufacturing (CAM) files.}
with the appropriate connector to plug straight into an instrument
(see Table~\ref{tbl:bom}). The board can be powered from a
\SI{5}{\volt} supply via the exposed header, or the USB ports on the
Arduino or ESP development board.

For RS232, we use the MAX202 \cite{max202}, which is a TTL-to-RS232
adapter. RS232 devices must be able to withstand a short-circuit to
ground, and incoming voltages of up to $\pm\SI{25}{\volt}$, which are
constraints the ESP controller cannot satisfy, but the MAX202 can. The
MAX202 additionally contains a charge pump doubler and inverter to
produce signals of $\pm\SI{10}{\volt}$ from the \SI{5}{\volt} power
supplied. We also use a custom PCB for this
module.\footnotemark[\value{footnote}] IEEE-488 and RS-232 are
industry standards that have been used in test instrumentation since
the 1960s for automation and control. Our system therefore opens up
the very attractive possibility of retrofitting old high-end equipment with
wireless monitoring and controlling capabilities.

\section{Publishing Data to Server}
The sensor modules connect to an MQTT service
\cite{Mosquitto, MQTT} to which they transmit data (see
\figref{fig:system}). This service runs on a networked computer
in the lab, which forms our data collection server. Data messages are sent
as strings from each sensor module to the service with the value and
units of each measurement. The messages are `published' to different `topics'
on the server, which differentiates each signal being monitored. Other
clients are then able to `subscribe' to the `topics', and receive the
measurements.

When first deployed, a sensor module initially starts up its own WiFi
network and awaits configuration from a web-browser (a mobile phone
can be used for this purpose).  This configuration specifies the SSID
and password for the WiFi network to which the data collection server
is connected, as well as the address of the server. Once configured,
the device's network disappears, and the module starts transmitting
data.

In order for the data to be useful for analysis, we must have a way of
storing it. We use the ``agent'' program Telegraf \cite{InfluxOther}
to collect the data messages and store them in a time-series database,
InfluxDB \cite{InfluxDB}. The data flow is shown in
\figref{fig:system}.

\subsection{Other Data Sources}

It is straightforward to send measurements to the message queuing
service from sources other than our sensor modules, as the
messages are transported in a simple string format, and the message
queue interface is an open standard with many implementations
available. Instruments that have USB, FireWire or network interfaces
do not lend themselves easily to interfacing with the ESP controllers, but
using suitable drivers for these instruments connected to a computer (e.g., a small Raspberry Pi single board computer),
one can upload measurements to the message queue component of the
monitoring system to take advantage of the analysis tools described in
the following sections.

With an appropriate front-end, events can also be published to the
server from direct user input via a web browser, e.g. from a cell
phone. Examples of relevant events include periodic servicing
maintenance tasks such as filter and lubricant replacement.
Recently we have included the logging of shared equipment
sterilization prompted by regulations put in place due to the COVID-19
pandemic. COVID-19 also introduced an acute requirement for contact
tracing of laboratory personnel. We therefore currently log the user
occupation of the laboratory, with a sign-in board built using
NodeRED, a tool we also use for real-time analysis as discussed
below.\footnote{The registration of users entering and leaving lab
  could also be conveniently achieved with an RFID proximity reader as
  a node in our monitoring system.}

\section{Front End / Action Analysis}

The setup we describe uses separate tools for real-time analysis, and
for interacting with historical data (i.e., data stored in the
InfluxDB database).  Details on the services and their configuration
can be found with our code \cite{OurProject}.

Here we explain the user-facing tools that we use.

\subsection{Visualisation}
To visualise the data stored in the InfluxDB database we use Grafana
\cite{grafana} which offers ``drag and drop'' construction of
monitoring dashboards and database queries. Grafana is accessed
via a web-browser, allowing users to view the
data remotely and from multiple computers simultaneously. It can also
be made publicly and globally accessible via the internet, as we have done
\cite{hoodoo}.

\begin{figure}
	\centering
	\includegraphics[width=\linewidth]{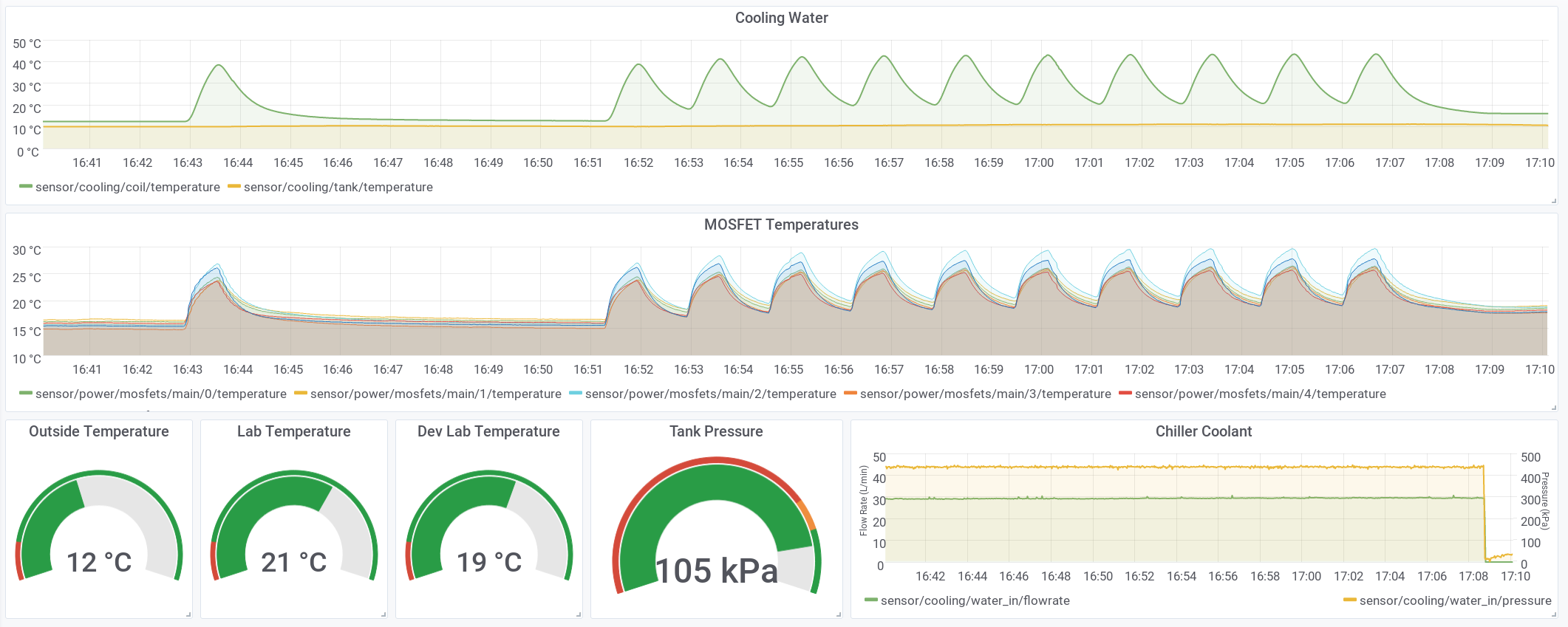}
	\caption{\label{fig:grafana}Screenshot of a monitoring dashboard
		built using Grafana. The top graph shows the temperature of our
		cooling water in a storage tank, and coming out of a electromagnet
		coil. The middle plot shows the temperature of MOSFETs used for
		controlling current. The bottom row has several gauge panels, with
		the green colour indicating normal operating points. The lower
		plot is the cooling water supply flow-rate and pressure.
		Reference \cite{hoodoo} provides a link to a live instance of
		our monitoring system.}
\end{figure}

Grafana provides a number of ways of visualising data as `panels' on a
dashboard. \figref{fig:grafana} shows a simple example using graph
and gauge panels. Grafana includes other panel types including bar
gauges, tables, and heat maps. With these tools, one can create ``control
panel'' type displays, or even ``digital checklists'', where the
operating state is indicated by colour, quickly highlighting anything
that is not operating normally. Grafana supports multiple dashboards,
and one can set the display to automatically change through them to
get a live overview of the system. Grafana also offers a simple way
for users to query and export data, avoiding having to interface with
the database directly.
\subsection{Real-time Analysis}

Our real-time analysis operates on data received from the message
queue server directly, so detects if there is a fault with the
experiment as soon as the data comes in. If a fault is detected, we
may either wish to avoid wasting time collecting faulty experimental
data, or we may try to act to correct the failure or minimise damage
caused by the failure. The last point is the realm of safety devices
(e.g. thermal fuses), and while it is not recommended to displace
hardware interlocks and fail-safes, real-time analysis opens up
interesting ways to augment these. An advantage of using the
monitoring system to perform these tasks is the large amount of
data available. For example, one may use data from multiple sensors
simultaneously, e.g. differential flow rates between source and return
lines can detect a leak.

For real-time analysis we use NodeRED \cite{NodeRED} which runs on the
data collection server and provides a web-based graphical and
JavaScript programming environment for reacting to messages from the
message queuing server. In NodeRED, messages are interpreted by a
stream processing pipeline. In our setup, anomalies are primarily detected by determining when directly
monitored quantities, or their combination, move outside set limits for normal
operating values. For example, as the filter in our cooling water inlet gets dirty, the flow rate of water will slowly reduce over time and once it falls below a lower limit the monitoring system flags this.  In addition, some parameters are compared against the average of recent values to detect sudden variations. An occurence of a leak or a  blockage in a branch of our multi-stranded cooling water supply may not cause the oveall flow rate to move outside the set limits, but will give rise to an abrupt change. More complex sensor fusion and feature extraction are possible within the NodeRED framework, but we
have found simple measures sufficient for detecting failures in our
ultracold atom machine.

We use NodeRED to perform some preventative actions, such as shutting
off our coolant pump if the system detects a leak or if the coolant
tank is low. We can also disable the power supplies if the
electromagnets being cooled get too warm or if the flow rate drops
off. It is worth again noting that, especially the last two of these,
should have dedicated hardware interlocks to act as a
fail-safe. Real-time analysis is not meant as a replacement for proper
safety design, but can take action to avoid the need to replace
thermal fuses. Additionally, it can notify the user that there has
been a fault, and where the fault occurred. In our setup, faults are
displayed on Grafana, a custom LED character display at the experiment
controls, and also sent out via Discord.

%\subsubsection*{Control Modules}
To take preventative actions we must have some devices capable of
effecting control over parts of the experiment. While we have focused
on the monitoring uses of our sensor modules, they are equally capable
of producing a control signal. Typical examples include controlling
relays, as shown in \figref{fig:system}, or triggering interlock
circuits. Hence the modules provide a way to react to changes in the
experiment, such as turn off a pump if a leak is detected.

The above control examples produce digital signals using the GPIO
pins. The ESP32 includes an \rhl{on-chip} digital-to-analog converter to
produce analog signals. The ESP8266 does not have an \rhl{on-chip}
converter, but one can be formed with pulse-width modulation
and an output filter. 
\section{Demonstrations}
\begin{figure}
	\centering
	\includegraphics[width=0.5\linewidth]{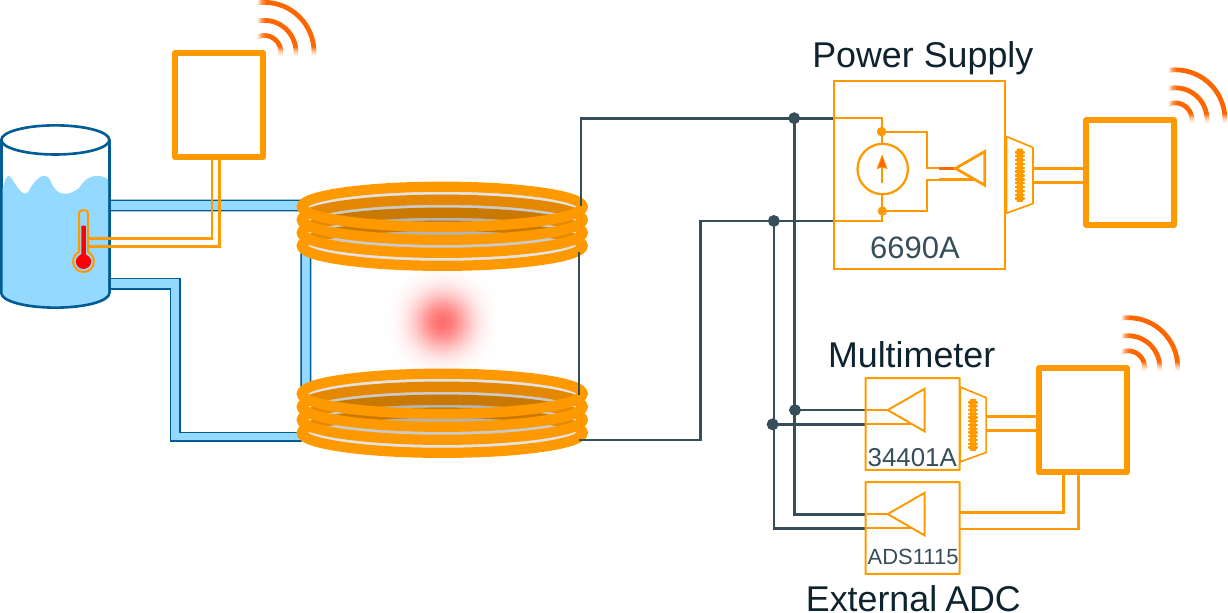}
	\caption{\label{fig:diagram} A test configuration for our system. A
		power supply sources a current to our quadrupole electromagnets,
		trapping a cloud of cold atoms in the resulting magnetic
		field. The coil voltage is monitored by the power supply itself, a
		GPIB compatible multi-meter, and an ADS1115 ADC. The
		electromagnets are cooled by a recirculating water system (chiller
		and pumps not pictured), with the temperature of the coolant tank
		monitored.}
\end{figure}
\begin{figure}[b!]
	\centering
	\includegraphics[width=0.5\linewidth]{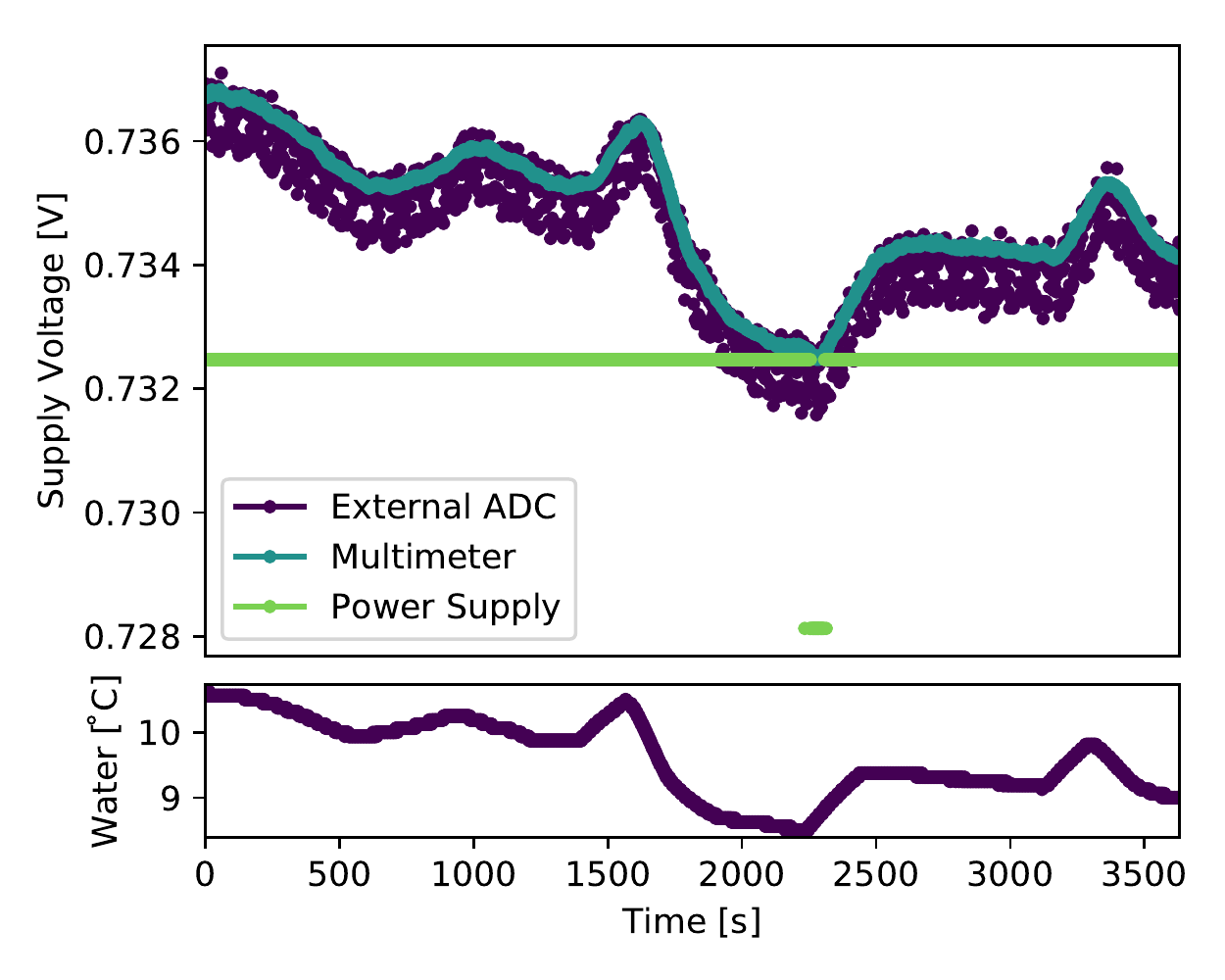}
	\caption{\label{fig:graph} Top panel: The voltage of the power
		supply for our the magnetic field coils of our quadrupole trap as
		measured in three different ways (see \figref{fig:diagram}). The
		power supply readout is not sufficiently sensitive to detect the
		drifts which are recorded by the external ADC and the 6.5 digit
		multi-meter. Bottom panel: The temperature of recirculating
		cooling water flowing in the hollow copper tubing of the coils as
		measured in the storage tank by a wireless module (see
		\figref{fig:diagram}).  }
\end{figure}
\subsection{Comparison of Data Sources}
As a demonstration of our monitoring system, we use the setup in
\figref{fig:diagram} to simultaneous log the voltage over a magnetic
field coil pair and the temperature of cooling water flowing in the
coils' hollow copper windings. The coils are driven at a constant
current by a commercial power supply (Keysight 6690A) and for
demonstration purposes we monitor the voltage drop in three ways: using
the module in \figref{fig:wiring_example}(e) the supply voltage is
monitored on one channel of an ADS1115 ADC , and using the wireless
GPIB modules of \figref{fig:photo}(a), we poll the power supply itself
as well as an external Keysight 34401A 6.5-digit multi-meter. The
temperature of the cooling water supplied to the coils is monitored
with the module in \figref{fig:wiring_example}(a).

Figure \ref{fig:graph} shows the data generated by the four detection
modules of \figref{fig:diagram} and collected by our monitoring system
over the course of an hour. Both the external ADC and the multi-meter
are sufficiently sensitive to pick up the drift in the power supply
voltage as it attempts to source a constant \SI{35}{\ampere}
current. The voltage fluctuations measured using both these modules
correlate highly with the temperature of the water cooling the coils,
suggesting that we are observing the change of resistance due to
temperature variation, with a coefficient of
\SI{51.2(1)}{\micro\ohm\per\kelvin}.

This example showcases the flexibility of our system. For simply
detecting a fault in the coil, the coarse readings from directly
polling the power supply itself via a wireless module would be
sufficient to detect an open or short circuit.
Using an ADC module, we can detect more subtle variations in
process parameters that might correlate from one end of the ultracold
atom machine to the other.
%for monitoring the
%health of our apparatus and track its evolution on a day-to-day basis.
Finally, using our wireless GPIB module we can achieve the
accuracy provided by a 6.5 digit commercial state-of-the-art voltmeter (Agilent 34401A) with a traceble calibration.

\subsection{Appending a spectrum analyzer}
\begin{figure}
  \centering
  \includegraphics[width=\linewidth]{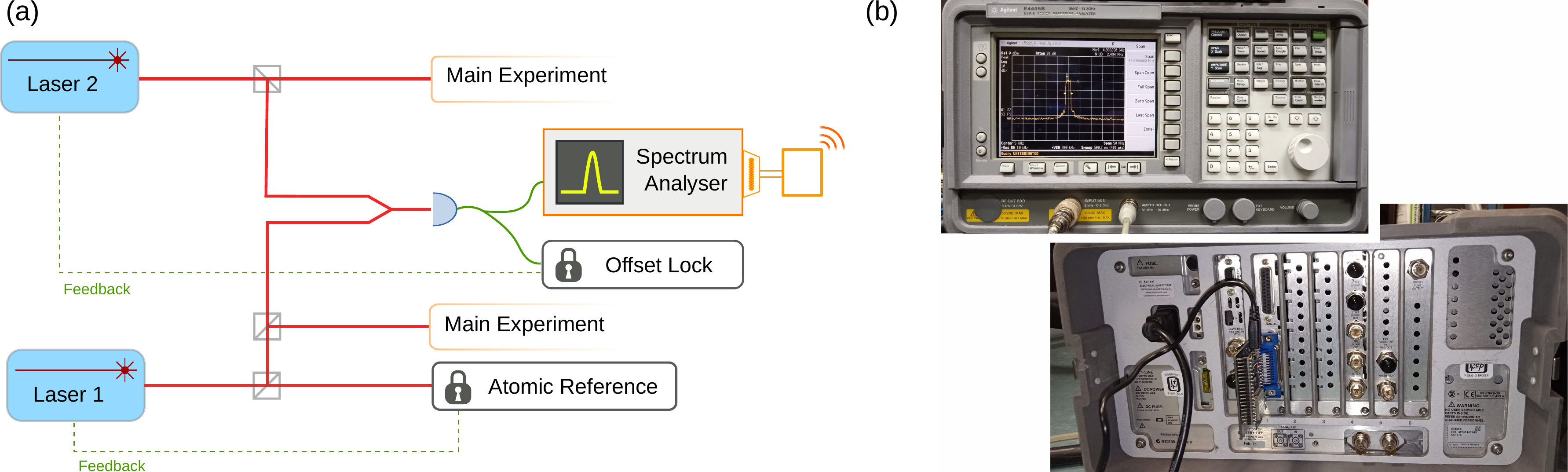}
  \caption{\label{fig:freq_diag} (a) Light from two lasers is
    combined to produce a beat-note on a photodiode. As the frequency
    of Laser 1 is locked to an atomic reference, the beat-note is used
    to lock the frequency of Laser 2. Additionally, we monitor the
    beat-note with a spectrum analyser controlled via GPIB from our
    sensor module. (b) The front and rear view of the spectrum
    analyser, showing showing the the beat-note signal on the display,
    and the compact wireless sensor module connected to the GPIB
    interface.}
\end{figure}
\begin{figure}[b!]
  \centering
  \includegraphics[width=0.6\linewidth]{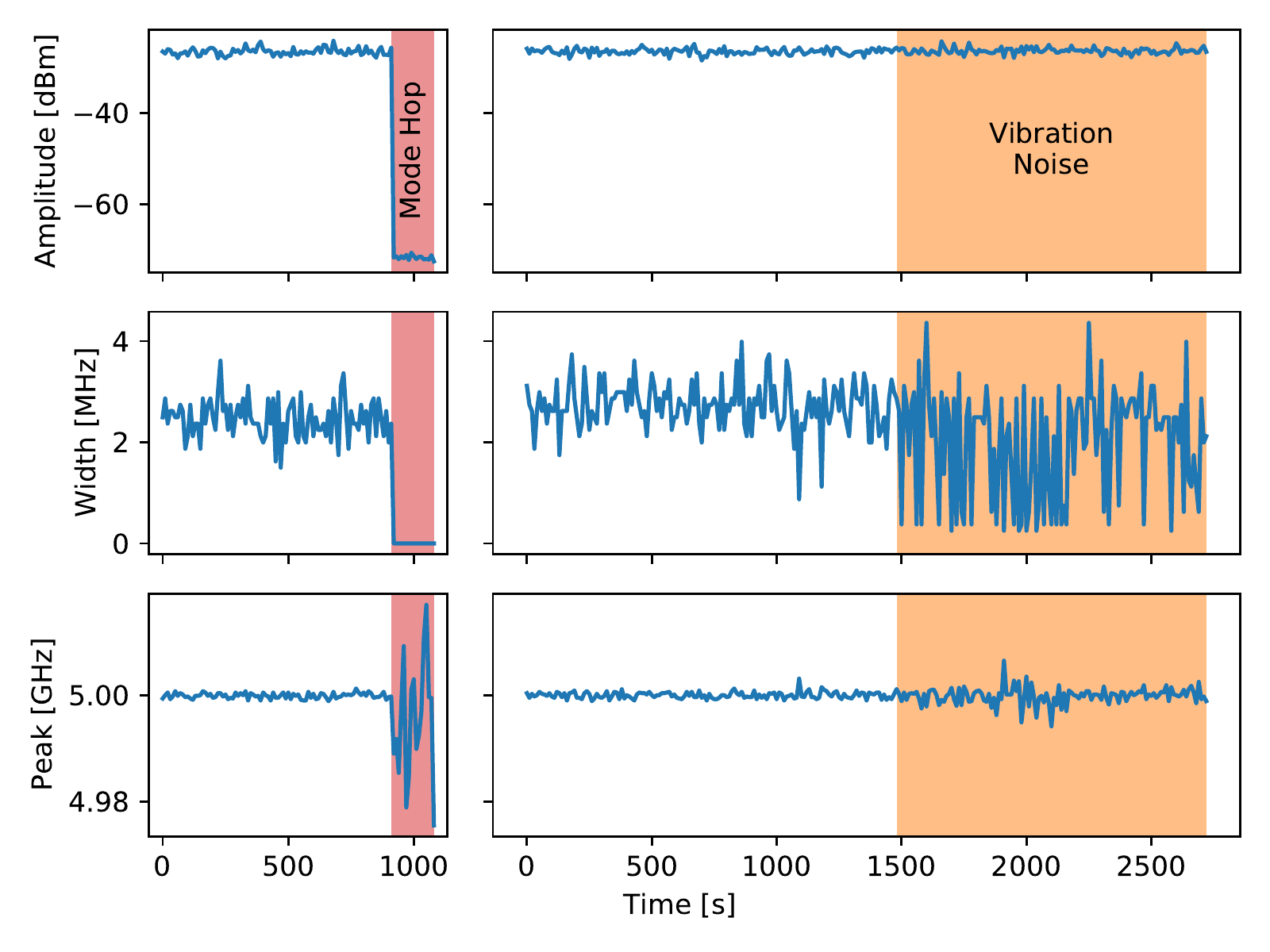}
  \caption{\label{fig:freqmon} The amplitude, linewidth, and centre
    frequency of the beat-note between two lasers, which are
    controlled such that the centre frequency is fixed at
    \SI{5}{\giga\hertz}. These observations show two areas of
    interest. In the red-shaded region, the laser has experienced a
    mode-hop, and the beat-note jumps out of range of the detector. In
    the orange-shaded region, an actively cooled RF source/amplifier
    is placed near the laser, coupling some vibration to the
    laser. The measured width becomes erratic as the beat-note moves
    around during a sweep of the spectrum analyser.}
\end{figure}
  Ultra-cold atom machines make extensive use of narrow-linewidth,
  tunable light to cool and probe atomic samples. The lasers producing
  this light are typically frequency locked to atomic references using
  absorption spectroscopy. With one laser tied to an atomic
  spectroscopic line, other slave lasers may be locked to this master
  through a so-called beat-note or offset lock \cite{Seishu2019,
    Appel2009}. Here, a phase detector compares the beat-note between
  the two lasers with a local oscillator. This offers an agile way of
  tuning the frequency of the slave laser, which can be offset by
  several GHz from the master. In our lab, we have found this
  particularly useful for dispersive optical probing of ultracold
  samples, where the probe light needs to be detuned far away from
  absorption resonances \cite{Sawyer2017}. Prior to the implementation
  of our condition monitoring system, we would manually assess the
  quality of a beat-note lock using an Agilent E4405A Spectrum
  analyser. This is an example of a high-end legacy instrument with
  exceptional performance but with communication limited to
  GPIB. Using our wireless GPIB interface described in section
  \ref{sec:serial}, we poll the amplitude, width and peak frequency of
  the beat-note between a master and slave laser as shown in
  \figref{fig:freq_diag}, and broadcast the values to the message
  queue of our monitoring system. Figure \ref{fig:freqmon} shows two
  periods of traced data, highlighting two classes of events. In the
  red region, the slave laser experienced a mode-hop, resulting in the
  complete disappearance of the beat-note within the span monitored by
  the spectrum analyser. In contrast to such an obvious event which is
  fatal to an atomic experiment, we can also observe more subtle effects:
  the orange shaded region shows a decrease in laser stability that
  arose from the introduction of an RF source and amplifier near the
  laser, which we have attributed to vibrations introduced by a
  cooling fan.

\section{Conclusion}

We have reported on a condition monitoring system targeted at the
research laboratory. Our platform operates with wireless sensor
modules distributed around the lab and feeding into a message queuing
service along with other data sources. The modules are all based on
wireless ESP microcontrollers that \hl{interface both with home-built
  sensors and with commercial laboratory equipment which are annexed
  by the monitoring system}. \rhl{This latter, distinctive feature can
  lease a new life to legacy, high-end equipment.} The present work
provides a vehicle for detecting acute equipment failures and for
providing information that quickly highlights the point of failure. It
also provides a framework for logging process parameters and operating
points of the day-to-day running of the experiments .

We have considered the specific example of an ultracold atom machine
and used this as a test bed, but our monitoring system is suitable for
a range of similarly complex experimental setups. The system is easily
deployed and detailed instructions are provided to do this
\cite{OurProject}, making collecting information about an experiment's
condition a small investment. The code we developed for this project
\cite{OurProject} is open-source, and uses third-party open-source
libraries and tools. We hope that our work this will stimulate further
development of this project.

\section*{Acknowledgment}
We thank Susanne Otto and Ryan Thomas for their
testing and feedback on this system. We acknowledge support from the Marsden Fund of New Zealand (contract UOO1923).

\bibliographystyle{elsarticle-num}
\bibliography{mattlab}

\appendix
\section{Supplementary material}\label{app}
\rhl{A supplementary note discussing the voltage tolerance of the ESP
  System-on-Chips and documenting the ADC non-linearity of the ESP32
  can be found online at (to be inserted by the editor). Code and
  detailed documentation for this project are available online
  \cite{OurProject}. This includes ready-to-use resources---sensor
  module firmware, service automation, and PCB designs for
  manufacture---and gives detailed documentation on how to replicate
  our monitoring system. }.

\end{document}